\documentclass[twocolumn,showpacs,aps,prl]{revtex4}

\usepackage{dcolumn}
\usepackage{bm}
\usepackage[pdftex]{graphicx}
\begin{document}

\title{STM images of sub-surface Mn atoms in GaAs: evidence of hybridization of surface and impurity states}

\author{J.-M. Jancu$^1$, J.-Ch. Girard$^1$, M. Nestoklon$^{1, 2}$, A. Lema\^{i}tre$^1$, F. Glas$^1$, Z.Z. Wang$^1$ and P. Voisin$^1$}
\affiliation{$^1$ CNRS-Laboratoire de Photonique et de Nanostructures,
  route de Nozay, F-91460, Marcoussis, France}

\affiliation{$^2$  Ioffe Physico-Technical Institut, Russian Academy of Science,
 194021 St Petersburg, Russia.}

\begin{abstract}
We prove that scanning tunneling microscopy (STM) images of sub-surface Mn atoms in GaAs are formed by hybridization of the
impurity state with intrinsic surface states. They cannot be interpreted in terms of bulk-impurity wavefunction imaging.
High atomic resolution images obtained using a low-temperature apparatus are compared with advanced, parameter-free
tight-binding simulations accounting for both the buckled (110) surface and vacuum electronic properties.
\end{abstract}

\date{\today}


\maketitle

The development of scanning tunneling microscopy has expended the applications of imaging to new areas of nanosciences and
engineering such as atom or molecule identification, manipulation and nanostructuring \cite{hla, braun, kitchen}. Both
spectroscopic and topological information is accessible, and quantum-size objects can be probed in real space with atomic
resolution. In particular, there is growing interest in the STM images of quantum dots \cite{marquez} and sub-surface
impurity states in semiconductors \cite{yakunin, mahieu,loth,marcz}. Although the well-accepted Tersoff and Hamann's theory
\cite{tersoff} simply relates the tunneling current to the local density of states (LDOS) at the tip position, the
interpretation of specific experiments on localized states is still a matter of debate: on the one hand, the interaction
between the surface and the quantum object under investigation must be examined; on the other hand, particularly in the case
of deep bound-states, the carrier escape toward the extended band states may play a role \cite{loth}. In the last few years,
acceptors in GaAs have attracted much attention because the shape of the images revealed strong and unexpected chemical
signatures: from triangles in the case of shallow acceptors like carbon \cite{loth} and zinc \cite{mahieu}, to asymmetric
butterfly for the deep acceptor manganese \cite{yakunin}. So far, theoretical work has concentrated mostly on comparison of
STM images with cross-sections of the bulk impurity wavefunction \cite{yakunin}. In this letter, we report on new
experimental data obtained with a low-temperature apparatus and compare these results with advanced TB calculations. We
first point that the STM image cannot reflect the LDOS of the impurity: indeed, empty-state STM images show only the
group-III elements whereas the neutral acceptor wavefunction is distributed over the different chemical species. In fact,
less than $40\%$ of the bulk impurity LDOS shows up in the STM image. Then we prove that the image actually results from the
hybridization between intrinsic surface states and the impurity state. A perturbative model is discussed and indicates that
a quantitative relation between STM images and bulk wave function is not straightforward. Supercell calculations based on
the $sp^3d^5s^*$ tight binding model and including the (110) surface as well as vacuum are performed and reproduce nicely
the experimental images.
\par
The sample used in this study was grown by molecular beam epitaxy on a GaAs(001) substrate at 420 C. It consists of two 40
nm-thick GaAs layers doped with $2. 10^{18}$ Mn cm$^{-3}$, embedded between 30 nm-thick GaAs conducting layers doped with
$1. 10^{19}$ Be cm$^{-3}$. The low temperature epitaxial growth was chosen in order to minimize the segregation of Mn atoms,
and the co-doping was needed because dilute Mn-doped GaAs is insulating below 77 K. Sample is cleaved in-situ and exposes in
a controlled manner the (110) or ($\overline{1}$10) surface. Figure 1a shows a mosaic of three overlapping constant-current
images measured at T=77K with sample-to-tip voltage $V_{st}$=+1.7V and current $I_{t}$=100pA, and Fig. 1b an
atomic-resolution image ($V_{st}$=1V, $I_{t}$=100pA) showing a few impurities. Strikingly, the atomic texture of Fig. 1b
shows only a rectangular 5.65 x 4 \AA$^2$ 2 D lattice corresponding to a single species sublattice, while the (110) surface
(Fig. 1d) exposes both Ga and As atoms. For n-doped GaAs, this feature was explained  \cite{feenstra,ebert} in the 1990's:
for positive $V_{st}$, electrons flow from the tip to empty surface states, consisting mostly of empty Ga dangling bonds,
whereas for negative $V_{st}$, electrons flow from occupied surface states (mostly As dangling bonds) to the tip. This is
why the image at positive $V_{st}$ shows the rectangular sublattice formed by Ga surface atoms and not the zig-zag chains of
Ga and As atoms along the [$\overline{1}$10] direction, characteristic of a (110) surface. The different shapes associated
with Be (triangle) and Mn (butterfly) dopants are observed simultaneously in Fig. 1a,b.

\begin{figure}[htp]
\centering{
\includegraphics[angle=0,height=0.6\linewidth, width=1\linewidth]{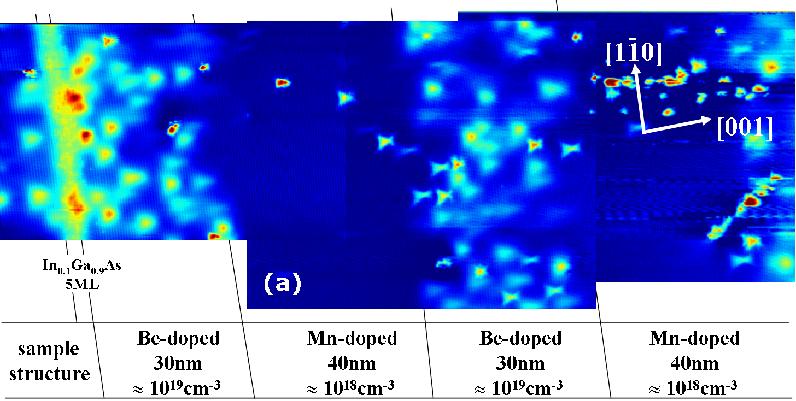}
\includegraphics[height=0.55\linewidth, width=1.\linewidth]{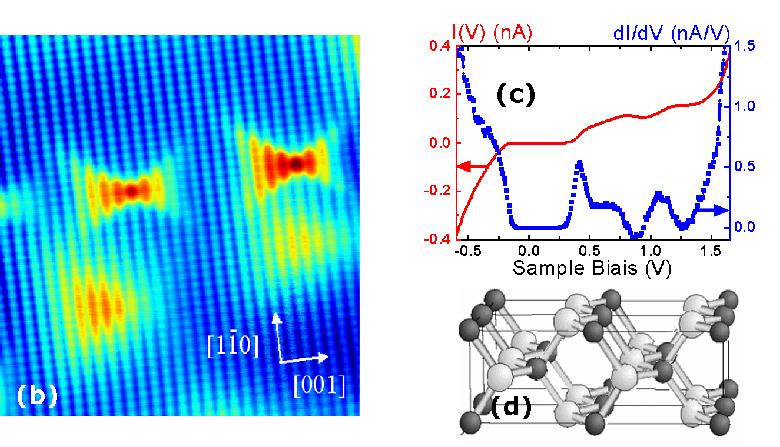}}
\vspace{-0.2cm}
\caption{(color online) (a) Mosaic of three overlapping constant-current images measured at T=77K, with sample-to-tip voltage Vst=+1.7V,
current It=100pA . The Be-doped and Mn-doped GaAs layers are clearly identified. The triangular-shaped and
butterfly-like images correspond to  Be and Mn dopants, respectively. (b) Atomic resolution image showing a few impurity
states. (c) Typical $I_{t}$($V_{st}$) and $dI_{t}$/$dV_{st}$ curves measured at fixed height in the center of a Mn impurity image. (d) Sketch of
atomic arrangement on a GaAs (110) surface}
\vspace{-0.5cm}

\end{figure}

Clearly, the texture of image within an impurity is essentially identical to that in the background. Conversely, the LDOS of
a neutral acceptor state in bulk GaAs must be spread over both types of atomic sites because it is built from zone center
valence states that are known to have 70 \% As and 30 \% Ga character. This can be observed in the calculations of
ref.\cite{yakunin}, and is further evidenced by the present $sp^3d^5s*$ TB approach. Mn is a deep acceptor whose binding
energy (113 meV) is governed by a strong central cell correction. In a TB frame, this local potential is associated with Mn
$d$-states and their hybridization  with $p$ states on neighboring As, a situation often handled perturbatively by shifting
in energy the on-site energy of a virtual group III element \cite{yakunin}and changing the nearest neighbor interactions. To
go beyond this approximation, we constructed transferable Slater-Koster parameters using an empirical spd nearest-neighbor
model with the nominal atomic structures: 4$s^24p^14d^05s^0$ for Ga, $3d^54s^24p^0$ for Mn, and 4$s^24p^34d^05s^0$ for As.
The parameters were optimized to fulfill the requirement of very good agreement with {\it ab-initio} calculations, keeping
unchanged the As one-site parameters between MnAs and GaAs (including a valence band offset of 100 meV) for different
prototype crystal structures (hexagonal and cubic). This procedure generates transferability allowing a precise modeling of
the local neighborhood of the impurity in the bulk. Electronic band structure calculations were performed by considering a
4096-atom supercell of zinc-blende GaAs (45 \AA $\times$ 45 \AA $\times$ 45 \AA ) in which one Ga atom is replaced by Mn.
The atomic coordinates where relaxed by minimizing the elastic energy using Keating's valence-force-field approach. A bound
state is found 90 meV above the valence-band edge, which is in fair agreement with experiment for the neutral acceptor state
of Mn\cite{chapman}. The following limitations of the present calculations are: i) Coulomb interaction between the remote
hole and Mn ion is not taken into account, which is reasonable since the binding energy is mainly due to the "central cell"
potential;  ii) the short-range $p-d$ exchange interaction giving a magnetic contribution to the binding energy of 25 meV
\cite{akb} is not considered. A 3 D plot of the resulting state and corresponding cross-sections in the (110) and
($\overline{1}$10) planes located 3 atomic planes away from the impurity center are shown in Fig. 2. The similar weights of
Ga and As in the wave function are obvious (see also left column in Fig. 4). Note that the cross-sections in (110) and
($\overline{1}$10) planes obey the rotoinversion symmetry about the impurity center, which is also observed in our measured
STM images.

\begin{figure}[t]
\centering{
\includegraphics[height=0.7\linewidth,width=1 \linewidth] {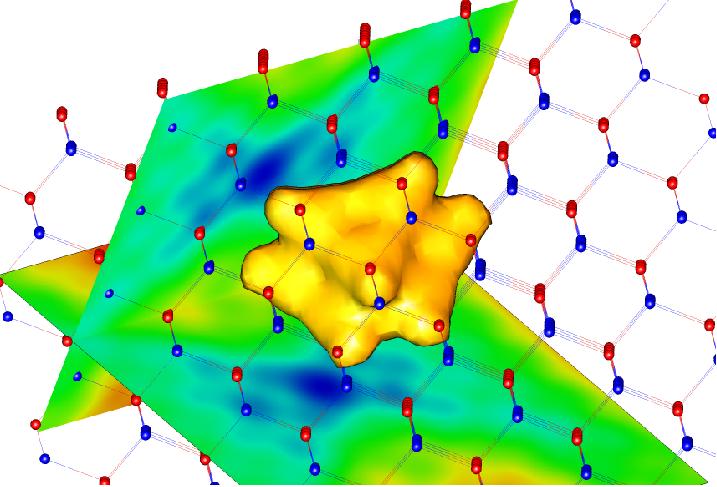}}
\caption{(color online) 3-D plot of the impurity bound state, and cross sections in the (110) and ($\overline{1}$10) planes situated 3 atomic planes
apart from the impurity center.}
\vspace{-0.5cm}
\end{figure}

\par
The strong difference between textures of STM image and cross-sections of a bulk impurity suggests that the former is
actually formed by hybridization of the impurity and surface states. Providing that the interaction is weak enough for a
perturbative approach to be valid, we can write the wave function of this mixed state as a linear combination:
\newcommand{\bra}[1]{\left\langle {#1} \right\vert}
\newcommand{\ket}[1]{\left\vert {#1} \right\rangle}
\newcommand{\matrel}[3]{\left\langle {#1} \left\vert {#2} \right\vert {#3} \right\rangle}
\newcommand{\braket}[2]{\left\langle {#1} \vert {#2} \right\rangle}
\newcommand{\intk}[1] {\int{ {#1} \; \mathrm{d}\bm{k}} }
\begin{equation}
\ket{\psi}=\alpha\ket{i} + \intk{\rho(\bm{k})\ket{s,\bm{k}}}
\end{equation}

of impurity wavefunction $\ket{i}$ and surface band wavefunction $\ket{s,\bm{k}}$. In first-order perturbation theory,
amplitude of the admixture of surface band is expressed as:
\begin{equation}\label{perturbation}
\rho(\bm{k})/\alpha = \left[ \frac{V(\bm{k}) }{E_i-E_s(\bm{k})} - I(\bm{k}) \right]
\end{equation}
where $E_i$ is the impurity level energy and $E_s(\bm{k})$ the surface band dispersion. $I(\bm{k}) = \braket{s,\bm{k}}{i}$
and $V(\bm{k})=\matrel{s,\bm{k}}{U_i}{i}$ (where $U_i(\bm{r})$ is the impurity potential)are the overlap and perturbation
integrals, respectively. $I(\bm{k})$ should be small compared to unity in any situation where the perturbative approach is
valid. Conversely,  evaluation of $V(\bm{k})$ requires precise knowledge of the impurity potential and detailed description
of the surface band structure. Equation 2 provides insight into the mixing between surface bands and impurity level, but
does not bring a quantitative information. To go beyond it is necessary to simulate the physical situation numerically. We
first prove that the present TB approach accounts fairly well for the surface states by calculating the electronic
properties of a superlattice where 40 monolayers of GaAs alternate with 3 nm of vacuum. Here vacuum is described in terms of
a zinc-blende crystal with TB parameters reproducing the dispersion of the free electron, and having a dielectric  constant
equal to unity. As discussed in ref. \cite{jancu98}, the $sp^3d^5s^*$ model is close enough to numerical completion to allow
for such folding of the free electron band structure. The rearrangement of surface atoms known as the buckling relaxation
which prevents the formation of surface states inside the gap \cite{ebert,engels} was taken into account in the calculation.
The band structure and LDOS of buckled GaAs(110) surface are found in excellent agreement with {\it ab initio} results, as
evidenced in Fig. 3 for the first empty dangling-bond state C3 \cite{ebert,engels}. The calculation closely reproduces the
general properties of C3 obtained from {\it ab initio} modeling in terms of energy, the strong Ga contribution to the
surface wave-function, and attenuation within the crystal and vacuum. The slow decay of surface states inside the crystal
certainly plays a major role in their strong interaction with sub-surface localized states.

\begin{figure}[t]
\centering{
\includegraphics[height=0.4\linewidth, width=0.4\linewidth] {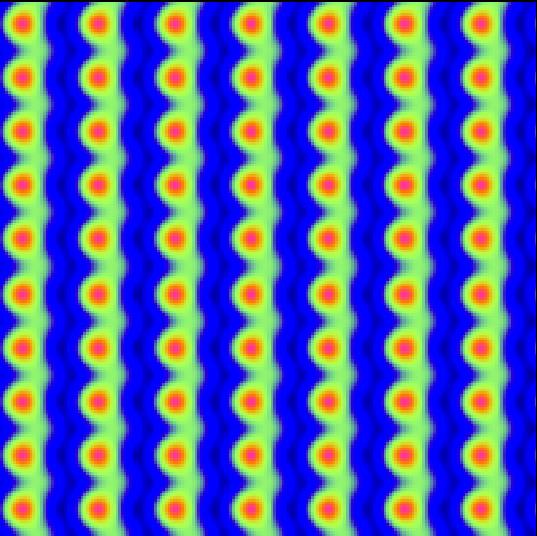}
\hspace{0.2cm}
\includegraphics[height=0.4\linewidth, width=0.4\linewidth] {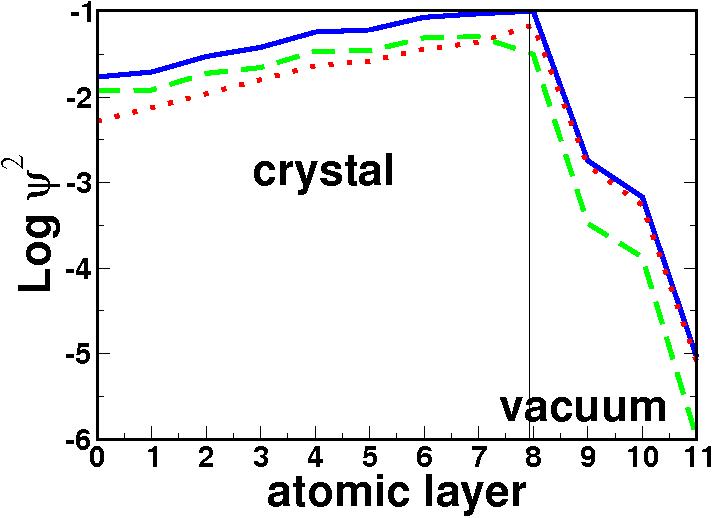}}
\caption{(color online) (a) LDOS plot of the surface state C3, at 0.5eV above the conduction band minimum. LDOS is calculated at $2 \AA$
above the surface. (b) Decays of C3 in crystal and vacuum along the [110] direction (solid line) and site-projected
anion (dashed line) and cation (dotted line) contributions.}
\vspace{-0.cm}
\end{figure}
\par
Next, a Mn dopant is added in the GaAs slab, in the nth plane below the surface. In Fig. 4, the cross-sections of the
corresponding states in the first plane of "vacuum atoms" (i.e. $2 \AA$ above the surface \cite{rem} are compared with
experimental results and cross-sections of the bulk impurity for n=3 to n=5. For these depths, the calculated acceptor
binding energy shows little change (from 90 to 85 meV). For n= 2 and n=1, the decrease of binding energy is more pronounced
while for the surface impurity, we find a large increase due to complete change of local environment. Discussion of n=0 to 2
is beyond the scope of this Letter. All the images have an horizontal mirror symmetry axis, but for even n, this axis
corresponds to a row of Ga atoms while for odd n, it goes through a row of As atoms.
\begin{figure}[b] \centering{
\includegraphics[height=1.\linewidth, width=1.\linewidth] {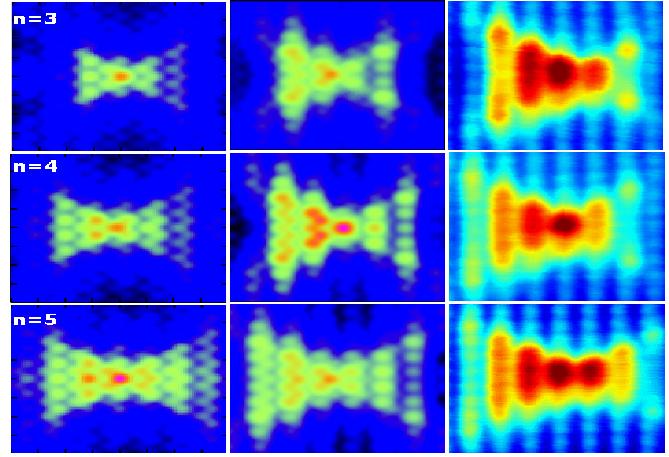}}
\caption{(color online) Bulk impurity cross section (BICS)(left), simulated (SSTM)(center) and experimental (right) STM images of a Mn
neutral acceptor located n monolayers (n=3 to 5) below the (110) surface. BICS is calculated in a 110 plane, n atomic planes
away from the impurity, and SSTM $2 \AA$ above the surface. Note that As rows are on the right side of Ga rows.}
\vspace{-0. cm}
\end{figure}
Major differences between the two types of theoretical images are clearly observed. The most striking discrepancy concerns
the respective weights of anion and cation sites: to be specific, we focus the discussion on the case of n = 4. The bulk
impurity cross-section (BICS) in that case is centered on a Ga site, and one easily checks that most of the bright spots
correspond to As sites. More precisely, the As sites dominate strongly the left part of the image, while Ga and As have
similar weights in the right part. Conversely, in the simulated STM (SSTM) image, the Ga sites clearly dominate and As show
up only in the nearest vertical rows on the left of the impurity. Similar conclusions are drawn for the n=3 and n=5 cases,
with the difference that corresponding images appear centered on an As site. The intensity distributions near the image
center also differ significantly, with a very sharp peak on odd-n BICS and a more distributed maximum in corresponding SSTM,
while the opposite holds for even-n images. Altogether, these features can be summarized by stating that hybridization
filters out the As component from BICS. This finding is by no way trivial: building on the perturbative model, one might
have expected that impurity state interacts preferentially with energetically closer valence-type surface states that have a
strong As component in vacuum \cite{ebert}. There are also more subtle and counter intuitive features: while the decay of
intensity with distance to center is roughly monotonic in BICS, one observes bright spots in SSTM and experimental images.
This is particularly obvious on the last visible vertical row on the right of the images. The distance between these bright
spots gives a direct and non-ambiguous measurement of the impurity depth: it is equal to n times the surface lattice spacing
along the [$\overline{1}$10] axis, or, equivalently, the number of weak spots between the two bright ones is equal to n-1.
It is also noteworthy that the global aspect ratio of the experimental images is better reproduced in SSTM than in BICS,
which indicates that part of the STM image asymmetry arises from the interaction with surface states. A small remaining
discrepancy can be observed as one more vertical row appears on the left of experimental images, compared to simulation. In
fact, this discrepancy is of the order of differences between similar experimental images and is likely due to uncontrolled
variations of impurity environment. Electric field due to tip induced band bending might for instance play a role. It should
also be noted that we compare constant current images with constant-height calculated LDOS, which certainly contributes to
small distortions of experimental $\it{vs}$ simulated images. Also, it is possible and would certainly be interesting to
include in the theoretical model Coulomb and p-d exchange interactions, but since the agreement is already within
experimental fluctuations, their influence on the STM images is likely limited and at most similar to that of uncontrolled
parameters. Finally, we insist that in the theoretical images of Figs. 3 and 4, the anions rows are on the right side of
cation rows. The agreement with observed STM images implies that it is also the case experimentally. From a crystallographic
point of view, this depends on an arbitrary convention regarding the identification of [110] $\it{vs}$ [$\overline{1}$10]
directions, and there are conflicting discussions in the literature \cite {zigzag,zigzag1} regarding this point. In our
experiment (see Fig. 1) the growth axis direction is unambiguously identified, and we can state firmly that for the cleavage
surface showing the asymmetrical butterfly with a stronger left wing, the anions are on the right side of cations. According
to commercial wafer specification, this is the (110) surface.
\par
We finally comment spectroscopic data on these images. $I_{t}(V_{st})$ curves at fixed height were recorded on many images,
and a typical spectrum is shown in Fig. 1c. These curves show a threshold whose position varies between 100 mV and 600 mV. A
clear correlation between threshold voltage and distance to the heavily p-doped layer is observed, suggesting that
variations of the tip-induced band bending is the main cause of threshold variation. Images of a given impurity show
essentially no dependence on $V_{st}$ until $V_{st}$ exceeds 1.5 V and $I_{t}$ starts integrating contributions from
tunneling to conduction band states. We point that STM measures basically a DC current, that can flow only if the tunneling
electron escapes out of the bound state at a rate faster than the injection rate \cite{loth1}. Since when changing $V_{st}$, one changes
the electrostatic environment of the impurity from hole depletion to hole accumulation, the very mechanism of electron
escape probably changes while the image is not affected. This strongly suggests that in our sample the voltage threshold
merely corresponds to a condition for the impurity to be effectively in an $A_{0}$ state, or for the electron to escape fast
enough. The detailed mechanisms of electron escape towards extended band states in STM spectroscopy certainly deserves more
attention, but in the present case it does not seem to influence the image itself.
\par
In conclusion, we have shown that STM images of sub-surface impurities are formed due to hybridization of impurity and
intrinsic surface states. These features are general and apply to other types of near-surface localized states like images
of shallow impurities (like Be in Fig. 1b)and quantum dots. Realistic tight-binding calculations including the surface allow
comparison with experiment at an unprecedented level of precision. The method can be extended to many different
nanometer-sized objects, up to a present limit of a few million atom supercells. More fundamentally, our analysis suggest
that the well understood bulk impurity state acts as an inner experimental probe of surface states and their extension in
the crystal.

Acknowledgement: The authors thank Dr. L. Largeau for valuable discussions and C. David for technical assistance.
Calculations were performed at the IDRIS-CNRS supercomputing center under project "CAPnano". One of us (JMJ) is supported by
the SANDIE NoE of the EC, and one of us (MN)by the "Dynasty" foundation.


\begin{thebibliography}{99}
\bibitem{rem} experimental conditions rather correspond to a tip to surface distance of $4 \AA$, but due to the fast decay of
surface states in the vacuum, we reach numerical underflow and loose image resolution at this distance. Note that C3 images
at $2 \AA$ and $4 \AA$ are quite similar.
\end{thebibliography}
\end{document}